\documentstyle[multicol,aps,prb,graphicx]{revtex}
\begin{document}
\draft
\title{Electron-Phonon Interactions in Correlated Systems:  Adiabatic Expansion
of Dynamical Mean Field Theory} 
\author{Andreas Deppeler and A. J. Millis}
\address{Center for Materials Theory, Department of Physics and Astronomy,
Rutgers University, Piscataway, New Jersey 08854} 
\date{\today} 
\maketitle
\begin{abstract}
We use the dynamical mean field theory to develop a systematic and computationally tractable method for studying electron-phonon interactions in systems with arbitrary electronic correlations. The method is formulated as an adiabatic expansion around the limit of static phonons. No specific electronic groundstate is assumed. We derive an effective low-frequency phonon action whose coefficients are static local correlation functions of the underlying electron system. We identify the correct expansion parameters. At a critical electron-phonon interaction strength the system undergoes a transition to a polaronic state. We determine the location of this polaronic instability in the presence of electron-electron interactions, doping, and quantum lattice fluctuations and present the formalism needed for study of the electron self-energy and effective mass.
\end{abstract}
\pacs{63.20.Kr, 71.27.+a, 74.25.Kc}
\begin{multicols}{2}
The development of a quantitative theory of electronic properties of ``correlated
materials'' (i.e.  those for which the local density approximation plus
Boltzmann transport theory is inadequate) is an important goal of materials
science.  An important feature of real materials is the electron-phonon
interaction.  The conventional theory of electron-phonon interactions in metals
is due to Migdal and Eliashberg (ME)\cite{m58} and is based on two assumptions:
that the underlying electronic state is well described by Landau's Fermi liquid
theory and that the typical phonon frequency $\omega_0$ is small compared to the
electronic Fermi energy $E_F$, so that an expansion is possible in the
``adiabatic parameter'' $\gamma = \omega_0/E_F$.  However, in many materials of
current interest electron correlations are strong, so a Fermi liquid 
description may not be appropriate and the relevant expansion parameter is 
unclear.

The introduction\cite{mv89} and more recent improvements\cite{gkkr96} 
of the dynamical mean field (DMF) method have opened an important avenue for progress,
by showing how a good approximation to the correlation physics can be obtained
from the solution of a quantum impurity problem plus a self-consistency
condition.  Unfortunately the straightforward inclusion of the electron-phonon
coupling in the DMF formalism presents a difficult technical problem: the
mismatch between the typical phonon frequency scale $\omega_0
\lesssim 0.1 {\rm eV}$ and electron energy scale $t \gtrsim 1 {\rm eV}$ renders
conventional numerical approaches to the impurity problem prohibitively
expensive, except in the ``anti-adiabatic'' limit \cite{fjs93,fzcj98}
$\gamma \gtrsim 1$ relevant to rather few materials.

In this paper we show how to turn this apparent difficulty to an advantage, by developing a systematic adiabatic expansion of the DMF formalism about the limit $\gamma = 0$ of static (classical) phonons which was shown to be easily tractable in Refs. \onlinecite{mms96a,cp99}. We derive an effective low-energy phonon action which, at leading order, reproduces ME theory, appropriately generalized to a ground state which is not necessarily a Fermi liquid. The simplifications inherent in the DMF theory allow us to go beyond leading nontrivial order and to calculate the effect of dynamic terms representing quantum lattice fluctuations. The vertices of the action are higher-order static local density correlation functions of the electron system in the absence of phonons. We identify the correctly renormalized expansion parameters and show that the expansion breaks down in the vicinity of a polaronic instability, occurring at a critical value of the electron-phonon coupling. We determine the value of this critical coupling using analytical and Quantum Monte Carlo methods. We also present the formalism needed to calculate the electron self-energy.

We write a general tight-binding based Hamiltonian as follows: $H = H_{\rm el} + H_{\rm ph} + H_{\rm el-ph}$. The electronic part is 
\begin{equation}
H_{\rm el} = -\sum_{ij\sigma}t_{i-j}
(c_{i\sigma}^{\dagger} c_{j\sigma} + c_{j\sigma}^{\dagger} c_{i\sigma}) - N \mu n + H_{\rm ee}.
\label{hel}
\end{equation} 
Here $H_{\rm ee}$ represents electron-electron interactions, not explicitly written.  
The operator $c_{i\sigma}^{\dagger}$ creates an electron with spin $\sigma$ on lattice site $i$.  The chemical potential is $\mu$ and the mean density is $n = (1/N)\sum_{i
\sigma} c_{i \sigma}^{\dagger} c_{i \sigma}$, where $N$ is the number of sites
in the lattice.

We model the phonons as quantum oscillators:
\begin{equation}
H_{\rm ph} = \frac{1}{2} \sum_{i}(M\dot{x}_{i}^2 + Kx_{i}^2).  
\end{equation} 
The operator $x_i$ measures the ionic displacement at site $i$, $M$ is the ion mass, and $K$ is the spring constant.  We have taken the oscillator frequency $\omega_0 = (K/M)^{1/2}$ to be dispersionless (Einstein model).  To apply the method to more realistic situations one should interpret $K$ and $M$ as averages over the appropriate phonon bands or use the extended DMF theory of Ref.  \onlinecite{mk00}.  Anharmonic terms\cite{fz01,fzj00} can be trivially added and will be seen to be generated by the electron-phonon interaction.

We take the electron-phonon interaction to be local:
\begin{equation}
H_{\rm el-ph} = g \sum_{i} x_{i} (n_i - n). \label{ephi}
\end{equation}
Here $n_{i} = \sum_{\sigma} c_{i\sigma}^{\dagger} c_{i\sigma}$, and $\langle x_i \rangle = 0$ is the equilibrium phonon displacement for a uniform electron distribution.

In DMF theory the properties of $H$ may be obtained from the solution of an impurity model specified by the action $S[c, \bar{c}, x, a] = S_0[x] + S_{\rm ee}[c, \bar{c}, a] + S_1[c, \bar{c}, x, a]$, with 
\begin{eqnarray}
S_0[x] & = & \frac{1}{2T} \sum_{k} x_{k} \left(K + M\omega_k^2 \right) x_{-k}, \label{s0} \\
S_1[c, \bar{c}, x, a] & = & - \sum_{n\,\sigma} \bar{c}_{n \sigma} c_{n \sigma} a_n + g \sum_{n k \sigma} \bar{c}_{n \sigma} c_{n + k,\sigma} x_k. \label{s1}
\end{eqnarray}
Here $S_{\rm ee}$ arises from $H_{\rm ee}$ and is not explicitly written. The impurity electron and phonon are represented by $c, \bar{c}$ and $x$, respectively, and have been Fourier transformed according to $x(\tau) = \sum_k \exp(-i\omega_k \tau) x_k$ etc.  Since the problem is local the fields depend on frequency but not on momentum.  We use the compact notation $x_k = x(i\omega_k)$, $c_n = c(i\omega_n)$, with bosonic (fermionic) Matsubara frequencies $\omega_k = 2k\pi T$ ($\omega_n = (2n+1)\pi T$) indexed by integers $k$ ($n$). The local Green function $G_{\rm loc}$ is calculated from the impurity partition function $Z[a] = \int [dc d\bar{c} dx] \exp -S[c, \bar{c}, x,a]$,
\begin{equation}
G_{\rm loc}[a]_n = \frac{\delta \ln Z[a]}{\delta a_n} \label{gloc} \equiv
\frac{1}{a_n - \Sigma[a]_n}, \label{gloc2}
\end{equation}
which defines the self-energy $\Sigma$. Notice that $Z$, $G_{\rm loc}$, and $\Sigma$ are functionals of the mean field function $a$ which is fixed by the condition that $G_{\rm loc}[a]$ agrees with the momentum integral of the full Green function, using the same self-energy, i.e.
\begin{equation}
G_{\rm loc}[a]_n = \int d\epsilon_k \frac{\rho(\epsilon_k)}{i\omega_n + \mu -
\Sigma[a]_n - \epsilon_k}, \label{gscc}
\end{equation}
which depends on the lattice density of states $\rho(\epsilon_k)$.

The foregoing is general.  To analyze the phonon problem it is convenient to formally integrate out the electron fields
\begin{equation}
\exp(-S[x,a]) = \int [dc d\bar{c}] \exp(-S[c,\bar{c},x,a]) \label{iout}
\end{equation}
and work with an effective phonon action $S[x, a] = S_0[x] +S_{\rm ee}[a] + S_1[x, a]$. We proceed by formally expanding $S_1$ about the values $\bar{x}$ which extremize $S$.  The crucial fact (of course already well-known to Migdal \cite{m58}) is that the characteristic phonon frequency scale is small.  Therefore, as we shall show, the frequency sums involving phonon fields are dominated by frequencies of order $\omega_0 = (K/M)^{1/2}$, thus $S_1$ may be evaluated by an expansion about the adiabatic limit.

For small $g$, $\bar{x} = 0$; this is the conventional metallic state with no lattice distortion.  As the coupling increases, eventually a state with $\bar{x} \neq 0$ becomes preferred. This corresponds to a polaronic instability at which the ground state is fundamentally reconstructed and has been extensively discussed \cite{mms96a} for models involving only electron-phonon interactions.  For the rest of this paper we assume $\bar{x} = 0$; i.e.  we expand about the undistorted ground state. We thus have:
\begin{eqnarray}
S_1[x, a] & = & -{\rm tr}\,\ln a \nonumber \\
&& - \sum_{n = 2} \frac{g^n}{n} \sum_{k_1,\ldots,k_n}\!\!\!'\;\Gamma_n[a]_{k_1, \ldots, k_n} x_{k_1} \cdots x_{k_n}, \label{actex}
\end{eqnarray}
where the prime denotes the restriction to $k_1 + \ldots + k_n = 0$. The $\Gamma_n[a]_{k_1, \ldots, k_n} \equiv \Gamma_n[a](i\omega_{k_1},\ldots,i\omega_{k_n})$ are the connected $n$-point local density correlation functions of the electronic action. Their explicit form depends on the non-phonon physics contained in $S_{\rm ee}$.  It is convenient to introduce a scale $t$ on which electronic properties vary (this could be the bandwidth or some interaction scale) and to define the parameters 
\begin{equation}
\gamma = \frac{(K/M)^{1/2}}{t} = \frac{\omega_0}{t}, \quad \lambda = \frac{g^2}{Kt}.
\label{glpar}
\end{equation}
We measure all frequencies and temperatures on the phonon frequency scale $\gamma t$,
writing $\tilde{\omega}_n = \omega_n/(\gamma t)$ and $\tilde{T} = T/(\gamma t)$,
and we rescale the phonon fields to $\tilde{x}_k \equiv
\tilde{x}(i\tilde{\omega}_k) = (K/T)^{1/2} x_k$ so the free action becomes
$S_0[\tilde{x}] = \frac{1}{2} \sum_k \tilde{x}_k(1 +
\tilde{\omega}_k^2)\tilde{x}_{-k}$, and the interaction part becomes
\begin{eqnarray}
S_1[\tilde{x}, a] & = & -{\rm tr}\,\ln a - \sum_{n = 2} \frac{\lambda^{n/2} (\tilde{T}\gamma)^{n/2 - 1}}{n} \nonumber \\
&& \sum_{k_1,\ldots,k_n}\!\!\!'\;\tilde{\Gamma}_n[a]_{k_1,\ldots,k_n} \tilde{x}_{k_1} 
\cdots \tilde{x}_{k_n}.
\label{iresc} 
\end{eqnarray} 
The parameter $\gamma$ now appears in the prefactor and in the frequency arguments of the rescaled vertex functions $\tilde{\Gamma}_n[a]_{k_1,\ldots,k_n} \equiv \tilde{\Gamma}_n[a](i\gamma\tilde{\omega}_{k_1}, \ldots,i\gamma\tilde{\omega}_{k_n})
= t^{n -1} T \Gamma_n[a](i\omega_{k_1},\ldots,i\omega_{k_n})$.  Provided
(as is usually the case) that the dominant contributions to $\tilde{\Gamma}_n$
come from frequencies of order $t$ we may evaluate the $\tilde{\Gamma}_n$ via a
low-frequency expansion and may for most purposes set the external frequencies
to zero and define coupling constants $\tilde{\Gamma}_n[a] :=
\tilde{\Gamma}_n[a]_{0,\ldots,0}$.  Notice that it is not necessary that the
electron Green function has no structure on the scale of $\omega_0$; merely that
the dominant contribution to $\tilde{\Gamma}_n$ comes from the scale of $t$.
With $S$ given as in Eq.  (\ref{iresc}), $\Sigma$, $G_{\rm loc}$, and $a$ may be
computed from Eqs. (\ref{gloc}), (\ref{gscc}) as formal power series in $\gamma$.

The complete action, up to terms of order $\gamma^{3/2}$, is
\begin{eqnarray}
&& S[\tilde{x}, a] = -{\rm tr}\,\ln a + \frac{1}{2} \sum_k \tilde{x}_k \tilde{D}^{-1}_k \tilde{x}_{-k}
\nonumber \\
&& - \frac{1}{3} \lambda^{3/2} \gamma^{1/2} \tilde{T}^{1/2}
\tilde{\Gamma}_3[a] \sum_{k_1, k_2} \tilde{x}_{k_1} \tilde{x}_{k_2}
\tilde{x}_{-k_1 - k_2} \nonumber \\
&& - \frac{1}{4} \lambda^2 \gamma \tilde{T}
\tilde{\Gamma}_4[a] \sum_{k_1, k_2, k_3} \tilde{x}_{k_1} \tilde{x}_{k_2}
\tilde{x}_{k_3} \tilde{x}_{-k_1 - k_2 - k_3}. \label{efalo}
\end{eqnarray}
Eq. (\ref{efalo}) can be regarded as an effective action on the low-energy scale $\gamma t$.  The phonon Green function is $\tilde{D}_k^{-1} = 1 + \tilde{\omega}_k^2 - \lambda \tilde{\Gamma}_2[a]_{k, -k} \approx 1 - \lambda/\lambda_c + \tilde{\omega}_k^2 + \lambda \gamma \alpha_p |\tilde{\omega}_k | + {\cal O}(\gamma^2)$, where $\lambda_c = \tilde{\Gamma}_2[a]^{-1}$ is a critical interaction strength and $\alpha_p$ is a damping parameter. The inclusion of static particle-hole diagrams $\tilde{\Gamma}_2$ in the phonon self energy reduces the phonon frequency scale but increases the electron-phonon interaction strength. The renormalized expansion parameters are
\begin{equation}
\bar{\gamma} = \gamma (1 - \lambda/\lambda_c)^{1/2}, \quad
\bar{\lambda} = \frac{\lambda}{1 - \lambda/\lambda_c}. \label{rexp}
\end{equation}
Physical properties $\sim \bar{\lambda}^m \bar{\gamma}^n$ (where $m$ and $n$ are the number of phonon lines and loops, respectively, in the corresponding Feynman diagram) diverge like $(1 - \lambda/\lambda_c)^{n/2 - m}$ at $\lambda \rightarrow \lambda_c$ (notice that $m \geq n$ always), and the uniform metallic state $\bar{x} = 0$ becomes unstable to local distortions.\cite{mms96a,bz98} Here we consider only $\lambda < \lambda_c$.

The coefficients $\tilde{\Gamma}_n$ in (\ref{efalo}) are higher order {\em static\/} susceptibilities of the interacting electron problem and may be computed with the techniques presently available. Here as an example we study the Holstein-Hubbard model with spin-$\frac{1}{2}$ electrons for which
\begin{equation}
S_{\rm ee} = U \int_{0}^{\beta} d\tau [n_{\uparrow}(\tau) - 1/2][n_{\downarrow}(\tau) - 1/2],
\end{equation}
where $n_{\sigma}(\tau) = \bar{c}_{\sigma}(\tau) c_{\sigma}(\tau)$. We specialize to a Bethe lattice for which $\rho(\epsilon_k) = (4t^2 - \epsilon_k^2)^{1/2}\theta(4t^2 - \epsilon_k^2)/(2\pi t^2)$ and the self-consistency equation (\ref{gscc}) becomes $a_n = i\omega_n + \mu - t^2 G_n$. Here we focus on $\tilde{\Gamma}_2$, but most considerations remain valid for $\tilde{\Gamma}_3$ and $\tilde{\Gamma}_4$. Combining Eqs. (\ref{iout}) and (\ref{iresc}) we can show that $\tilde{\Gamma}_2 = t T \langle (n - \langle n \rangle_{\rm ee})^2 \rangle_{\rm ee}$, the static local density-density correlation function, where the average $\langle \cdots \rangle_{\rm ee}$ is taken with respect to $S_{\rm ee}$. All calculations will be done in the paramagnetic (PM) state.

In the weakly interacting limit $u := U/(2t) \ll 1$ we use standard perturbation theory. To second order in $u$ we can replace the mean-field function $a$ entering $\tilde{\Gamma}_2[a]$ by its noninteracting form $a_0 = i\omega_n + \mu - t^2 a_0^{-1}$. We find
\begin{eqnarray}
&& \tilde{\Gamma}_2[a](u, \beta, \mu) = \tilde{\Gamma}_2[a_0](0, \beta, \mu) - u \tilde{\Gamma}_2^2[a_0](0, \beta, \mu) \nonumber \\
&& + u^2 \left(\frac{1}{2}\tilde{\Gamma}_2^3[a_0](0, \beta, \mu) + C_2(\beta, \mu) \right) + {\cal O}(u^3), \label{ptu}
\end{eqnarray}
where $\tilde{\Gamma}_2[a_0](0, \beta, \mu) = -(8/\pi) \int_{-1}^1 dx f(x - \mu) \sqrt{1-x^2} x$ is the noninteracting particle-hole bubble and $C_2(\beta, \mu)$ is given as a sum of integrals which can be evaluated numerically. We set $\beta = 2t/T$ and measure $\mu$ on the scale $2t$. Details will be given in a separate publication.

In the strongly interacting limit $u \gg 1$ we use Hartree-Fock (HF) theory\cite{rzk92} which approximates the local Green function by $G_n = \frac{1}{2}[(a_n - U/2)^{-1} + (a_n + U/2)^{-1}]$. The resulting cubic equation for $G_n$ may be evaluated numerically and the result used to calculate $\tilde{\Gamma}_2$. At half filling we find $\tilde{\Gamma}_2 \approx \frac{1}{4} u^{-3}$ for $u\rightarrow \infty$.

In order to study finite doping and the crossover from weak to strong electron-electron interactions we use the Quantum Monte Carlo (QMC) method as described in Ref. \onlinecite{gkkr96}. The imaginary time interval is divided into $L = \beta/\Delta\tau$ time slices. At each time slice an Ising spin is introduced to decouple the quartic $U$ interaction. The partition function and all observables can then be sampled stochastically over all $2^L$ auxiliary spin configurations. Computer time grows with $L^3$. The QMC algorithm has to be iterated self-consistently [using Eq. (\ref{gscc})] until a convergent solution for $a$ is obtained. We found 15-20 iterations to be sufficient. The starting guess for $a$ was computed using finite-temperature Iterated Perturbation Theory.\cite{gkkr96}

\begin{figure}
\includegraphics[width=8cm]{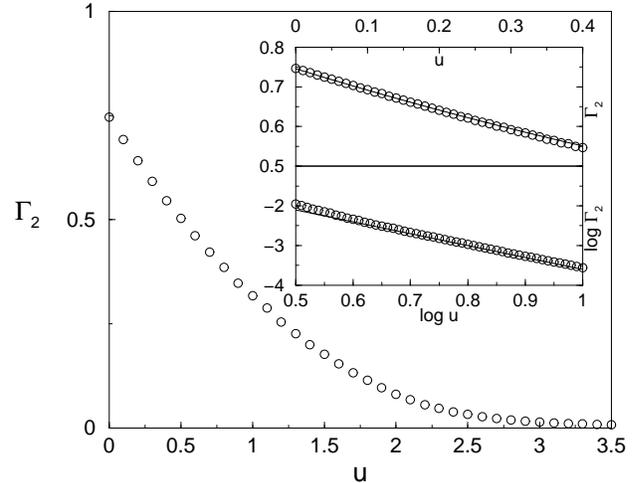}
\caption{Inverse critical coupling $\tilde{\Gamma}_2 = \lambda_c^{-1}$ as a function of electron-electron interaction $u = U/(2t)$, at half filling and $\beta = 6$. Circles: $L = 32$ QMC results. Lines: weak-coupling perturbation theory (upper inset), strong-coupling Hartree-Fock theory (lower inset; notice log-log scale).}  \label{g2u}
\end{figure}

Fig. \ref{g2u} shows the results of QMC simulations of $\tilde{\Gamma}_2$ as a function of $u$. Physical parameters are $\beta = 6$ and $\mu = 0$. The main graph shows the crossover from weak to strong coupling. In the insets we compare QMC (circles) to analytical results (lines), with excellent agreement. Upper inset: weak-coupling perturbation theory as discussed above, with $\tilde{\Gamma}_2[a_0](0, 6, 0) = 0.745702$ and $C_2(6,0) = -0.032789$. Lower inset: $\log \tilde{\Gamma}_2$ vs. $\log u$ which converges to the asymptotic solution $\log \tilde{\Gamma}_2 = -2\log 2 - 3\log u$ found above.

\begin{figure}[tb]
\includegraphics[width=8cm]{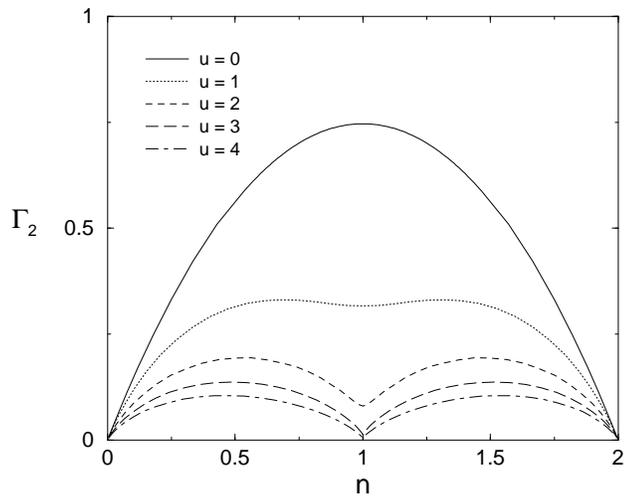}
\caption{QMC simulation of inverse critical coupling $\tilde{\Gamma}_2 = \lambda_c^{-1}$ as a function of electron density $n$, for various values of electron-electron interaction $u = U/(2t)$ and $\beta = 6$.}  \label{g2d}
\end{figure}

In Fig. \ref{g2d} we plot $\tilde{\Gamma}_2$ as a function of electron density $n$, for $\beta = 6$ and various $u$, computed by QMC with $L = 32$. For $u = 0$ single-particle density fluctuations decrease when the system is doped away from half filling, as expected from the Pauli principle. At $u = \infty$, density fluctuations are forbidden at half filling, so $\tilde{\Gamma}_2(n = 1) = 0$. The minimum at $n = 1$ develops as pairs (for $n \gtrsim 1$) or holes (for $n \lesssim 1$) become the dominant density fluctuations at low doping.

In general we expect $\lambda_c$ also to depend on lattice vibrations, i.e. non-zero $\gamma$. Integrating out cubic and quartic terms in Eq. (\ref{efalo}) leads to an effective quadratic action $S[\tilde{x}, a] = \frac{1}{2}\sum_k \tilde{x}_k (1 - \lambda/\lambda_c(\gamma) + \tilde{\omega}_k^2) \tilde{x}_{-k}$ where $\lambda_c^{-1}(\gamma) = \tilde{\Gamma}_2 + \frac{3}{4} \bar{\lambda} \bar{\gamma} \tilde{\Gamma}_4 + {\cal O}(\gamma^2)$ at half filling and $T = 0$. The vertex $\tilde{\Gamma}_4[a_0] = -16(1 - \mu^2)^{3/2}(1 - 6\mu^2)/(15\pi)$ is negative at $u = 0$. A $u \rightarrow \infty$ HF calculation gives $\tilde{\Gamma}_4 \sim -u^{-3}$ and no sign change. Thus $\tilde{\Gamma}_2$ (classical phonon self-energy) and $\tilde{\Gamma}_4$ (quantum lattice fluctuations) have competing effects: the former increases the coupling $\lambda \rightarrow \bar{\lambda}$ while the latter decreases $\bar{\lambda}$ by renormalizing $\lambda_c \rightarrow \lambda_c(\gamma)$. For superconductivity, a similar competition between ``phonon dressing'' and ``vertex corrections'' was observed in Ref. \onlinecite{mfn98}.

We finally consider the electron self-energy $\Sigma = \Sigma^{\rm ee} + \Sigma^{\rm ph} = \delta \phi/\delta G$, where the Luttinger-Ward functional $\phi[G]$ is the sum of all vacuum-to-vacuum skeleton diagrams. The electron part $\Sigma^{\rm ee}$ comes from $\phi$ arising from $S_{\rm ee}$, the phonon part $\Sigma^{\rm ph}$ from $S_1$ which may be expanded in powers of $\gamma$ as in Eq. (\ref{efalo}). To leading order we find
\begin{equation}
\Sigma^{\rm ph}_{n \sigma} = -\frac{\lambda}{2} \sum_k D_k \frac{\delta \tilde{\Gamma}_2[a]_{k,-k}}{\delta G_{n\sigma}} + {\cal O}(\gamma^2).
\end{equation}
In general $\Sigma^{\rm ph}$ is of order $\gamma$ and its frequency dependence is on the scale of $t$ so that it may be neglected compared to either the bare frequency dependence or $\Sigma^{\rm ee}$. However, if $\delta \tilde{\Gamma}/\delta G$ is singular at low frequencies, as in Fermi liquids, then $\partial \Sigma/\partial \omega$ may be of order unity. To investigate the possibility of such a singularity we note that
\begin{eqnarray}
\tilde{\Gamma}_2[a]_{k,-k} & = & -t T \sum_{n \sigma} G_{n \sigma} G_{n + k,\sigma} \times \nonumber \\
&& \left(1 + t T \sum_{n^{\prime} \sigma^{\prime}} \Lambda_{n n^{\prime} k}^{\sigma \sigma^{\prime}} G_{n^{\prime} \sigma^{\prime}} G_{n^{\prime} + k,\sigma^{\prime}} \right),
\end{eqnarray}
where the vertex $\Lambda$ is given in terms of the particle-hole irreducible vertex $\Lambda^I$ via
\begin{displaymath}
\Lambda_{n n^{\prime} k}^{\sigma \sigma^{\prime}} = \Lambda_{n n^{\prime} k}^{I \; \sigma \sigma^{\prime}} + t T  \sum_{n^{\prime\prime} \sigma^{\prime\prime}} \Lambda_{n n^{\prime\prime} k}^{I\;\sigma \sigma^{\prime\prime}} G_{n^{\prime\prime} \sigma^{\prime\prime}} G_{n^{\prime\prime} + k,\sigma^{\prime\prime}} \Lambda_{n^{\prime\prime} n^{\prime} k}^{\sigma^{\prime\prime} \sigma^{\prime}}.
\end{displaymath}
Following the usual Fermi liquid arguments we observe that $\Lambda^I$ is a smooth function of its arguments, so the required singular behavior can only occur if we differentiate on one of the explicit $G$ factors, leading to
\begin{equation}
\Sigma_{n\sigma}^{\rm ph} = \lambda t T \sum_k D_k G_{n + k,\sigma} \Lambda_{k\sigma}^2 + \Sigma_{n\sigma}^{\rm ph\,reg}
\end{equation}
with $\Lambda_{k\sigma} = 1 + tT\sum_{n^{\prime} \sigma^{\prime}} \Lambda_{0 n^{\prime} 0}^{\sigma \sigma^{\prime}} G_{n^{\prime} \sigma^{\prime}} G_{n^{\prime} + k,\sigma^{\prime}}$ and $\Sigma^{\rm ph\,reg}$ a function which varies on the scale of $t$ or $U$. If the ground state is a Fermi liquid then $G_{n\sigma} = -i\pi\, {\rm sign}(\omega_n) \rho(\mu) + G^{\rm inc}$ and we obtain for the phonon contribution to the mass enhancement
\begin{equation}
\left. \frac{m^{\ast}}{m} \right|_{\rm ph} = 1 + \bar{\lambda} t \rho(\mu) \Lambda_{0\,\sigma}^2.
\end{equation}
We have calculated $\Lambda$ and find it decreases rapidly as correlations increase, so the density-coupled electron-phonon interaction is ``turned off'' as the Mott insulator is approached. Details will be presented elsewhere.

To summarize, we have combined a small-$\gamma$ expansion with DMF theory to obtain a general formalism for studying electron-phonon effects in correlated electron systems. We have identified a local polaronic instability at $\lambda \rightarrow \lambda_c$ where physical properties diverge. We have shown that $\lambda_c$ can be increased by electron-electron interactions and quantum fluctuations and decreased by doping. We have obtained formal expressions for the one-phonon electron self-energy and effective mass. Future papers will apply the method to studies of conductivity, isotope effects, and dynamical consequences of electron-phonon interactions near a Mott transition.

We thank S. Blawid and R. L. Greene for useful discussions.  We acknowledge NSF Grant No. DMR00081075 and the University of Maryland-Rutgers MRSEC for support.  

\end{multicols}

\begin{references}
\bibitem{m58} A.  B.  Migdal, Sov.  Phys. JETP {\bf 7}, 996 (1958); G.  M.  Eliashberg, Sov.  Phys.  JETP {\bf 11}, 696 (1960) 
\bibitem{mv89} W.  Metzner and D.  Vollhardt, Phys.  Rev.  Lett.  {\bf 62}, 324 (1989) 
\bibitem{gkkr96} A.  Georges, G.  Kotliar, W.  Krauth, and M. J.  Rozenberg, Rev.  Mod.  Phys.  {\bf 68}, 13 (1996) 
\bibitem{fjs93} J.  K. Freericks, M.  Jarrell, D.  J.  Scalapino, Phys.  Rev.  B {\bf 48}, 6302 (1993)
\bibitem{fzcj98} J.  K.  Freericks {\em et al.}, Phys.  Rev.  B {\bf 58}, 11613 (1998) 
\bibitem{mms96a} A.  J.  Millis, R.  Mueller, and B.  I.  Shraiman, Phys. Rev.  B {\bf 54}, 5389 (1996) and  Phys.  Rev.  B {\bf 54}, 5405 (1996)
\bibitem{cp99} S.  Ciuchi and F.  de Pasquale, Phys.  Rev.  B {\bf 59}, 5431 (1999)
\bibitem{mk00} Y. Motome and G.  Kotliar, Phys.  Rev.  B {\bf 62}, 12800 (2000)
\bibitem{fz01} J. K. Freericks and V. Zlati\'{c}, Phys. Rev. B {\bf 64}, 073109 (2001)
\bibitem{fzj00} J. K. Freericks, V. Zlati\'{c}, M. Jarrell, Phys. Rev. B {\bf 61}, R838 (2000)
\bibitem{bz98} P. Benedetti and R. Zeyher, Phys. Rev. B {\bf 58}, 14320 (1998)
\bibitem{rzk92} M. Rozenberg, X. Y. Zhang, and G. Kotliar, Phys. Rev. Lett. {\bf 69}, 1236 (1992)
\bibitem{mfn98} P.  Miller, J.  K. Freericks, and E.  J.  Nicol, Phys.  Rev.  B {\bf 58}, 14498 (1998)
\end{references}
\end{document}